\documentclass[ALICE,manyauthors]{cernphprep}

\newcommand{\pp}{p\kern-0.05em p}
\newcommand{\ppbar}{\mathrm{p}\kern-0.05em \overline{\mathrm{p}}}

\newcommand{\GeV}{\ensuremath{\mathrm{GeV}\kern-0.05em}}
\newcommand{\GeVc}{\ensuremath{\mathrm{GeV}\kern-0.05em/\kern-0.02em c}}

\usepackage[comma,square,numbers,sort&compress]{natbib}
\usepackage{hyperref}
\usepackage{lineno}
\usepackage{amsmath}
\usepackage{bigints}  
\usepackage{afterpage}  
\usepackage{multirow}
\usepackage{xcolor}
\usepackage[Symbol]{upgreek}
\usepackage[T1]{fontenc}
\usepackage{mathtools}
\usepackage{siunitx}
\usepackage{soul}

\usepackage{orcidlink}

\usepackage{xspace}
\usepackage{color}
\usepackage{xcolor}
\usepackage{rotating}
\definecolor{light-gray}{gray}{0.8}

\newcommand{\MeVc}{\ensuremath{{\rm MeV/}c}\xspace}

\newcommand{\pt}{\ensuremath{{\it p}_{\rm T}}\xspace}
\newcommand{\Rl}{\ensuremath{{\it R}_{\rm L}}\xspace}

\newcommand{\sqrtspp}{$\sqrt{s}~=~13~{\rm TeV}$}

\newcommand{\ptrange}[2]{{#1 $<$ \pt $<$ #2  \GeVc}\xspace}

\newcommand{\etaless}[1]{\ensuremath{\left|\eta\right| < #1}\xspace}

\newcommand{\yless}[1]{\ensuremath{\left|y\right| < #1}\xspace}

\newcommand{\ptD}{\ensuremath{{\it p}_{\rm T, D^{0}}}\xspace}

\newcommand{\Dzero}{\ensuremath{{\rm D}^{0}}\xspace}
\newcommand{\Dzerobar}{\ensuremath{\overline{{\rm D}^{0}}}\xspace}

\newcommand{\cmnt}[1]{}

\newcommand{\jetptrange}[2]{{#1 $<$ \ptjet $<$ #2  \GeVc}\xspace}

\newcommand{\ptjet}{\ensuremath{{\it p}_{\rm T}^{\rm ch.\ jet}}\xspace}

\begin{document}%

\begin{titlepage}
\PHyear{2025}
\PHnumber{082}      
\PHdate{01 April}  

\title{Energy--energy correlators in charm-tagged jets in proton--proton collisions at $\sqrt{s} = 13$~TeV} 
\ShortTitle{EEC in charm-tagged jets in pp at $\sqrt{s} = 13$~TeV}   

\Collaboration{ALICE Collaboration\thanks{See Appendix~\ref{app:collab} for the list of collaboration members}}
\ShortAuthor{ALICE Collaboration} 

\begin{abstract}
In this letter, we present the first measurement of the energy--energy correlator (EEC) in charm-tagged jets from proton--proton (pp) collisions at $\sqrt{s} = 13$~TeV. EECs probe the structure of QCD radiation, providing a unique test of mass-dependent effects in parton showers involving a charm quark and offering a distinct view into non-perturbative phenomena, including the hadronization process. The EEC is measured for charm-tagged jets and flavor-untagged (inclusive) jets with transverse momenta of {\mbox{\ptrange{10}{30}}}, where charm-quark mass effects are significant. We observe a significant suppression of the EEC amplitude in charm jets compared to inclusive ones, consistent with the expected suppression of radiation from massive quarks --  a fundamental prediction of QCD. Despite the significant amplitude difference, the observed peak positions of the charm and inclusive-jet EEC are similar, indicating a complex interplay between Casimir factor (differentiating quark and gluon jets), and quark-mass (dead-cone) effects in the QCD parton shower as well as subsequent hadronization effects. Comparisons with next-to-leading order calculations and various Monte Carlo event generators reveal the sensitivity of this observable to both mass effects in the parton shower and hadronization process. These results provide new constraints on theoretical models of heavy-quark jets and offer insights into the parton-to-hadron transition in QCD.

\end{abstract}
\end{titlepage}

\setcounter{page}{2}

The production of heavy quarks (charm and beauty) in hadronic collisions occurs during the hard-scattering processes in the early stages of the collision. Since their masses are much greater than the characteristic scale of quantum chromodynamics (QCD), $\Lambda_{\rm QCD}$, heavy quark production is well-described by perturbative calculations~\cite{ALICE:2011aa, Abelev:2012vra, Aaij:2013mga, ALICE:2019cbr}. 
Partons including heavy quarks which are produced by large momentum transfer 
 undergo a cascade of gluon radiation and subsequent quark--antiquark pair creation (parton shower) until their virtuality decreases to the scale of $\Lambda_{\rm QCD}$. At this point, the partons hadronize, collectively forming jets. Studying these jets provides direct information about the energy and momentum of the parton initiating the shower. Additionally, the internal structure of a jet probes the evolution from the initial hard-scattered parton to the final-state hadrons. 
 
 QCD predicts that the radiation pattern from a parton depends on its mass and color charge. In particular, the parton shower is sensitive to the different Casimir color factors of quarks and gluons~\cite{ALTARELLI1977298}, as well as to the quark mass, the latter resulting in the so-called dead-cone effect~\cite{Yu_L_Dokshitzer_1991}. This effect predicts a suppression of gluon radiation emitted by a heavy quark of mass $m_{\rm q}$ and energy $E_{\rm q}$, within a cone of angular size $m_{\rm q}/E_{\rm q}$ around the radiator. Due to these flavor effects, gluon-initiated showers are broader than quark-initiated ones, while heavy-quark-initiated showers fragment harder than those initiated by light quarks or gluons. While the splitting functions for light quarks and gluons have been extensively measured~\cite{CMS:2017qlm, ALICE:2021obz, ALICE:EEC, Aad_2020, chang2021inclusivejetcrosssection}, achieving an experimentally clean separation between gluon-initiated and quark-initiated jets remains a significant challenge.  

A number of heavy-flavor (HF) jet and jet substructure measurements have been previously reported by ALICE~\cite{ALICE:deadcone, 2023, Acharya_2023, Acharya_2024, 2019, 2022}, ATLAS~\cite{ATLAS:jetshape, Aad_2012}, CMS~\cite{PhysRevD.98.092014, Sirunyan:2019rfa, Sirunyan_2020} and LHCb~\cite{Aaij_2022, Aaij_2017, lhcbcollaboration2024measurementspsi2schic13872production}. In this Letter, we explore the Casimir factor (differentiating quark and gluon jets), and quark-mass (dead-cone) effects in the QCD parton shower and put new constraints on the hadronization mechanism through the first measurement of the {\textit {energy--energy correlator}} (EEC) for heavy-quark initiated jets. 
The EEC observable is derived from quantum field theory and is infrared and collinear safe, allowing for precise theoretical calculations~\cite{Hofman:2008ar, Chen_2020}. Experimentally, the EEC is defined as the energy-weighted cross section to produce a particle pair with a given opening angle ($R_{{\rm L}}$), where the energy weight is assigned as the product of the energy carried by each particle, normalized by the hard scale of the physics process. By focusing on jets rather than the entire event, we probe the detailed structure of QCD radiation in charm-induced showers~\cite{Basham:1979gh,Basham:1978zq,Basham:1978bw,Basham:1977iq,Chen:2020vvp}.  The EEC observable in this paper, $\Sigma_{\rm EEC}(R_{\rm L})$, is defined as the two-particle correlation function and is given by
\begin{equation}
    \Sigma_{\rm EEC}(R_{\rm L}) = \frac{1}{N_{\rm jet}\Delta}\int_{R_{\rm L}-\frac{1}{2}\Delta}^{R_{\rm L}+\frac{1}{2}\Delta} \sum_{\rm jets} \sum_{i, j}\left(\frac{p_{{\rm T}, i}p_{{\rm T}, j}}{p_{\rm T, jet}^{2}}\right)\delta(R_{\rm L}'-R_{{\rm L}, ij}) {{\rm d}R_{\rm L}'},
\end{equation}
where the jet transverse momentum (${p_{\rm T, jet}}$) is used as the hard-scale reference in the calculation of the energy weight. The summation runs over all final-state particle pairs ($i$, $j$) inside each jet \footnote{Both pairs ($i, j$) and ($j, i$) are considered. For example, a pair of particle 1 and particle 2 is counted twice, (1, 2) and (2, 1).}. The angular distance between each pair in the $\eta-\varphi$ plane is $R_{{\rm L}, ij} = \sqrt{\Delta\varphi_{ ij}^{2}+\Delta\eta_{ ij}^{2}}$, where $\varphi$ and $\eta$ are the azimuthal angle and pseudorapidity, respectively. The angular interval size is denoted by \( \Delta \), and \( N_{\rm jet} \) represents the total number of jets. EECs probe the dynamics of jet formation, governed by QCD evolution equations~\cite{Gribov:1972ri, Dokshitzer:1977sg,MARCHESINI199549, ALTARELLI1977298} at different angular scales, $R_{\rm L}$~\cite{ALICE:EEC, PhysRevLett.130.051901}. Energy flow in the large $R_{\rm L}$ region is dominated by early partonic splittings, while the small $R_{\rm L}$ region corresponds to later radiation and is dominated by the dynamics of final-state hadrons~\cite{PhysRevLett.130.051901}. This separation between the predominantly perturbative and non-perturbative regions, as predicted by~\cite{Neill:2022lqx,PhysRevLett.130.051901}, is demonstrated by the recently published measurement of inclusive-jet EEC in pp collisions by ALICE~\cite{ALICE:EEC} and CMS~\cite{CMS:2024mlf}. The intermediate to low $R_{\rm L}$ region is where confinement occurs. Comparing the EEC in this region between inclusive jets and HF jets will provide valuable insights into hadronization mechanisms, which remain not fully understood.

Recent theoretical calculations have shown that for jets, the mass of heavy quarks distinctly imprints a characteristic angular scale in the EEC. This is evident from the turnover observed precisely at the angular scale defined by the heavy-quark mass, providing direct access to mass effects such as the dead-cone effect~\cite{craft2022beautifulcharmingenergycorrelators}. Utilizing the ALICE detector's excellent tracking and particle-identification capabilities, jets can be tagged with fully reconstructed heavy-flavor hadrons. This paper presents the first measurement of $\Sigma_{\rm EEC}(R_{\rm L})$ for charm-tagged jets, identified by the presence of a \Dzero meson among the jet constituents, in pp collisions at \sqrtspp. In addition, we measure $\Sigma_{\rm EEC}(R_{\rm L})$ for flavor-untagged (inclusive) jets in the same kinematic region as the charm-tagged jets, both with and without a cut on leading-track within the jets. These results provide a new test of perturbative QCD calculations, including mass effects, particularly at low jet \pt, where the charm-quark mass represents a significant fraction of the jet's energy. By comparing the EEC of charm-tagged jets with that of inclusive (gluon-dominated) jets, we can directly study flavor-dependent effects in parton showers and hadronization. Additionally, the results are compared with state-of-the-art Monte Carlo (MC) models, providing new constraints on MC event generators and hadronization models~\cite{Skands_2014, Bellm:2015jjp}. This measurement also provides a baseline for future studies in heavy-ion collisions.
 
\newline
The results presented here are obtained from pp collisions at $\sqrt{s}=13$ TeV recorded by ALICE during the LHC Run 2 data taking period between  2016 and 2018. Minimum bias events are selected by triggering on coincidence hits in the arrays of scintillator detectors~\cite{ALICE:2013axi} located at forward pseudorapidity $ 2.8 < \eta < 5.1$ (V0A) and backward pseudorapidity $-3.7 < \eta < -1.7$ (V0C), in order to minimize beam--gas background. Additionally, events are required to have a primary vertex within $\pm 10$ cm from the nominal interaction point to ensure uniform detector acceptance. Events with more than one reconstructed vertex are rejected to suppress pileup. A total of $1.5 \times 10^{9}$ minimum bias events are analyzed, corresponding to an integrated luminosity of 25.81 $\pm$ 0.43 nb$^{-1}$~\cite{ppXsec}. Charged particles are reconstructed using the central-barrel tracking system. This includes the Inner Tracking System (ITS)~\cite{ALICE:2010tia}, the Time Projection Chamber (TPC)~\cite{Alme:2010ke}, and the Time-Of-Flight detector (TOF)~\cite{CERN-LHCC-2000-012,Cortese:545834}. The central-barrel detectors are placed in a 0.5 T magnetic field which is oriented along the beam direction. The ITS provides precise determination of primary and secondary vertices, as well as high-resolution tracking of charged particles. The TPC ensures tracking of charged particles and contributes to particle identification (PID) via specific energy loss (d$E$/d$x$). The TOF detector, combined with the TPC energy-loss measurements, enables efficient PID across a broad momentum range by providing PID from time-of-flight measurements. ALICE has a high reconstruction efficiency of charged particles within the transverse-momentum range of $0.15 < \pt <100 \ \GeVc$. The single track reconstruction efficiency rises from $\sim 60\%$ at 0.15~\GeVc~and plateaus at $\sim 80\%$ for tracks with  $p_{\rm T} >$ 0.4~\GeVc. The track transverse-momentum resolution, $\sigma_{p_{\rm T}}/p_{\rm T}$, is approximately $ 1\%$ at {\mbox{$\pt=1$ \GeVc}}, increasing to about $ 4\%$ at $\pt=50$ \GeVc. The angular resolution is approximately $ 1$ mrad at $\pt=1$ \GeVc~ and decreases to less than $0.6$ mrad for $\pt>5$ \GeVc. A more detailed description of the ALICE detector and its performances can be found in~\cite{ALICE:2014sbx,ALICE:2008ngc}. 

\newline
In this work, heavy-quark initiated jets are identified by the presence of a prompt D$^0$ meson, which originates directly from a charm quark, among their constituents. Prior to jet finding, D$^0$-meson candidates are reconstructed via the decay mode ${\rm D^{0} \rightarrow K^{-} \pi^{+}}$  (with branching ratio \mbox {{${\rm 3.93} \pm {\rm 0.04}\%$}}~\cite{dmeson_decayL}) within the rapidity range \yless0.8. The transverse momentum of candidates selected for the analysis is constrained to the $ 5 < \ptD< 30$ \GeVc~ interval. The \Dzero meson and its antiparticle (\Dzerobar) are treated equivalently and are referred to as \Dzero in the following, unless otherwise specified. The combinatorial background from pairs not originating from D$^0$ decays is suppressed by applying geometrical selections on the displaced decay topology, along with particle identification of the hadron daughters. The cuts used for the track and \Dzero-candidate selections are based on those in Ref.~\cite{2023}. The decay daughters of the \Dzero-meson candidates are then replaced by their 4-momentum vector sum (i.e. the D$^0$-meson 4-momentum vector). This is done to mitigate the cases where the angle between the decay daughters is larger than the jet radius, as this would degrade the jet-energy resolution. D$^0$-meson tagged charged-particle jets are then reconstructed from the \Dzero candidates plus charged particles with the anti-$k_\mathrm{T}$ algorithm~\cite{Cacciari_2008} implemented in the FastJet software package~\cite{Cacciari:2011ma} with a resolution parameter $R = 0.4$ and $E$-scheme recombination. Except for the \Dzero-meson candidate, the pion mass is assumed for all jet constituents. The tracks are selected with the same criteria as reported in Ref.~\cite{ALICE:EEC}, which assure uniform acceptance in $\eta$ and $\varphi$. The 
charged particles are required to have {\mbox {\pt ~> 150 \MeVc}} and \etaless 0.9. Charged-particle jets containing fully reconstructed \Dzero candidates are selected in the transverse-momentum range $10 < \ptjet < 30$ \GeVc, and the jet axis is required to be within the fiducial volume of the TPC, $|\eta_{\text{jet}}| < 0.9 -R$. After the jet selection, the EEC observable is calculated for each jet. The $\Sigma_{\rm EEC}(R_{\rm L})$ contribution from background D$^0$ candidates that survived the selection criteria is removed using a sideband subtraction procedure, following the same procedure as in Ref.~\cite{2023}, yielding a true \Dzero-tagged jet EEC in intervals of \ptD.

In the next step, the D$^0$-tagged jet EEC distributions are corrected for four effects. First, a correction for the efficiency of \Dzero-tagged jet reconstruction within the detector acceptance is applied. The efficiency is estimated using MC simulations performed with the PYTHIA~8.244~\cite{Bierlich:2022pfr} with the Monash tune~\cite{Skands_2014} and a GEANT3 model~\cite{Brun:1119728} of the ALICE detector. The EEC distributions are corrected for the efficiency in ${\it p}_{\rm T, D^{0}}$ intervals. The efficiency-corrected EEC distribution in a given jet-${\it p}_{\rm T}$ interval is obtained by summing over all ${\it p}_{\rm T,D^{0}}$ intervals.  Second, jets tagged with a D$^0$ meson coming from a B-meson decay (non-prompt D$^0$) are subtracted. This background is estimated using POWHEG~\cite{Alioli2010} + PYTHIA~8~\cite{Skands_2014} + EvtGen~\cite{LANGE2001152} simulations. POWHEG is a Next-to-Leading Order (NLO) event generator which can more accurately reproduce the measured non-prompt-D$^0$ cross section compared to standalone PYTHIA~8 simulations. PYTHIA~8 is used for the showering and hadronization processes, while EvtGen is used to simulate the B-meson decay kinematics. The non-prompt \Dzero-tagged jet particle-level distribution is first scaled by the integrated luminosity and the branching ratio. It is then corrected for the prompt reconstruction efficiency (to match the non-prompt \Dzero-tagged jet EEC fraction in the uncorrected data) as well as the non-prompt reconstruction efficiency. Additionally, a bin-by-bin correction factor (explained later) is applied to account for detector effects, ensuring the distribution is converted to the detector level before subtracting the non-prompt contribution. The non-prompt contribution ranges from 10\% to 15\% (20\% to 25\%) from the lowest to the highest \(R_{\rm L}\) intervals in the 10--15 (15--30) \(\text{GeV}/c\) jet-\pt intervals. Third, the EEC distributions are corrected for detector effects, such as track-momentum resolution, angular resolution, and both single-track and pair inefficiencies. The track-momentum resolution affects the energy weights and \ptjet. The number of reconstructed pairs is influenced not only by single-track inefficiency, which results in missing reconstructed tracks, but also by mis-reconstructed tracks due to the track-merging effect~\cite{Acharya_2018}. 

As each track pair has a unique energy weight, unfolding to correct for detector effects is complicated. A 3-D (\ptjet, \Rl, energy weight) unfolding is required, which is currently not possible due to the limited available statistics. However, the excellent momentum and position resolution of the ALICE tracker minimizes bin-to-bin migration effects in the $R_{\rm L}$ distribution. Hence, in this analysis, instead of unfolding, a bin-by-bin correction is utilized where the correction factors are obtained from simulation by comparing the generator-level and corresponding detector-level information for each \ptjet and $R_{\rm L}$ bin~\cite{ALICE:EEC}. Finally, the resonance decay (${\rm c} \rightarrow {\rm D}^{*+} \rightarrow {\rm D}^{0} \pi^{+}$ and charge conjugate) forms a second peak in the low \Rl region (hadronic), which is not linked to the charm shower and therefore needs to be removed. The correction factor is determined by taking the ratio of the EEC distribution in PYTHIA~8 with the ${\rm D}^{*+}$ decay turned off to that with the ${\rm D}^{*+}$ decay turned on. Applying this correction factor to the detector-effect corrected EEC distribution yields the fully corrected charm-tagged jet EEC distribution. The inclusive-jet EEC baseline measurement follows the same jet-reconstruction procedure and detector-effects correction as the published results in Ref.~\cite{ALICE:EEC}.
\newline 
Systematic uncertainties related to the reconstruction of \Dzero-meson candidates are assessed by varying the topological and particle-identification criteria used for \Dzero-meson identification. The systematic uncertainty due to signal extraction is evaluated by varying the bandwidths of the signal and background, and performing multiple trials of the invariant-mass fit while varying the fitting settings.
The systematic uncertainty associated with the B-meson feed-down correction is estimated by varying input parameters in the POWHEG simulations~\cite{Cacciari_2012}. These parameters include the b-quark mass, renormalization ($\mu_{\rm R}$) and factorization ($\mu_{\rm F}$) scales in the perturbative calculation, and the parton distribution function (PDF). The uncertainty on single-track reconstruction efficiency is estimated to be 3\%~\cite{Acharya_2019}. To assess its impact, 3\% of detector-level tracks in the MC simulation are randomly rejected, and the analysis is repeated with the updated reconstruction efficiency and correction factor. The systematic uncertainty due to limited angular resolution is evaluated by applying an additional track-pair selection criterion ($\Delta \eta > 0.004$), and comparing the corrected result to the default case where there is no selection cut on track pairs. 

The dominant sources of systematic uncertainty are the \Dzero-meson selection efficiency and the track-pair reconstruction. The uncertainty on the \Dzero-meson selection efficiency ranges from 11\% to 6\% (5\% to 3\%) depending on \(R_{\rm L}\) intervals for the 10--15 (15--30) \(\text{GeV}/c\) jet-\pt ranges, while the tracking-efficiency uncertainty varies from 11\% to 8\% depending on \(R_{\rm L}\) intervals for both jet-\pt ranges. The total systematic uncertainty is 17\% to 5\% (17\% to 8\%) from the lowest to the highest $R_{\rm L}$ intervals in 10--15 (15--30) \GeVc~  jet-\pt intervals, obtained by summing the different contributions in quadrature, considering they are uncorrelated.

The sources of systematic uncertainty and their estimation for the inclusive-jet EEC follow the same procedure as the published results in Ref.~\cite{ALICE:EEC}. The total systematic uncertainty is $\sim$5\% in both jet-\pt intervals, for inclusive-jet EEC measurements both with and without the leading-track \pt selection. 
\begin{figure}[!ht]
    \centering
     \includegraphics[width=1\textwidth]{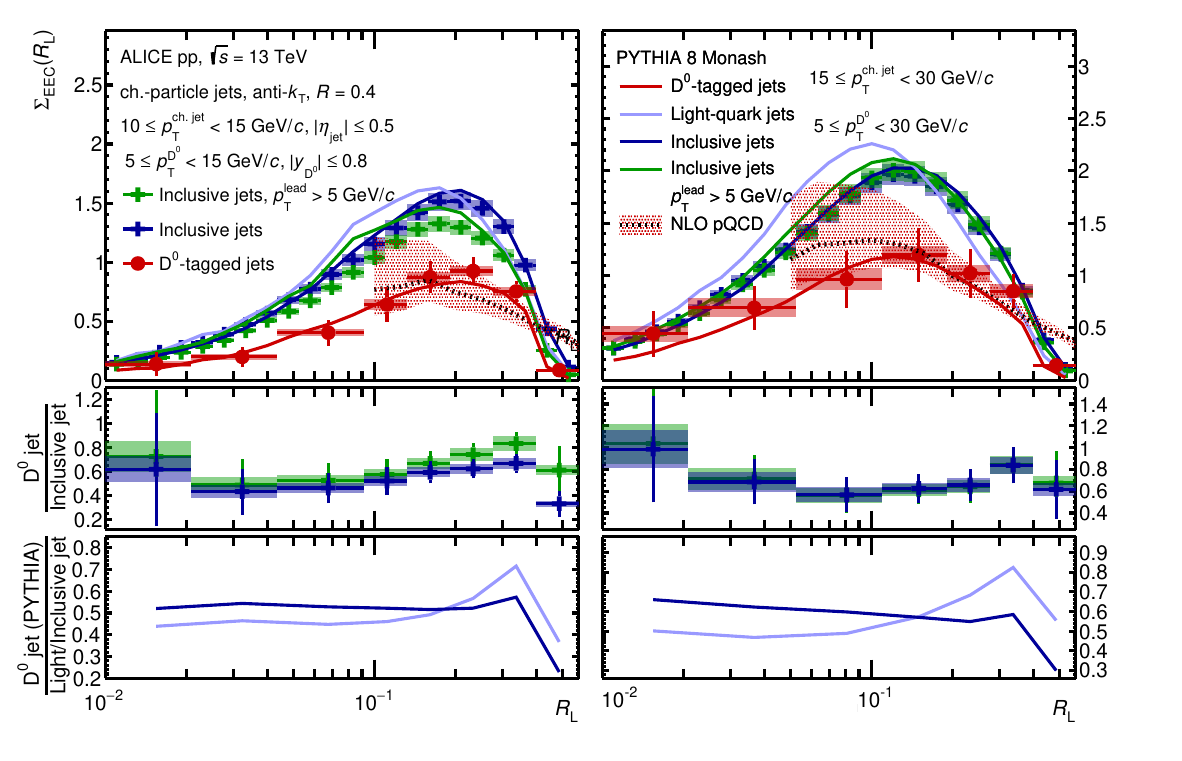}
   
    \caption{\Dzero-tagged jet (red) EEC distribution compared to inclusive jets with no leading track \pt selection (blue) and inclusive jets with a 5 \GeVc~leading-track \pt selection (green), to PYTHIA~8 simulations, and to pQCD predictions~\cite{craft2022beautifulcharmingenergycorrelators} in the \jetptrange{10}{15} (left) and \jetptrange{15}{30} (right) jet-\pt intervals. The middle panel shows the ratio of \Dzero-tagged jets to inclusive jets in the data. The bottom panel compares the ratio of \Dzero-tagged jets to inclusive jets with the ratio of \Dzero-tagged jets to light-quark jets in PYTHIA~8. The error bars show the statistical error, and the boxes represent the systematic error.}
    \label{fig:EECs_comp_pythia}
\end{figure}

Figure~\ref{fig:EECs_comp_pythia} shows the measured $\Sigma_{\rm EEC}(R_{\rm L})$ observable for charm-tagged (reconstructed \Dzero-meson) jets, compared to inclusive jets (mixture of gluon- and quark-initiated jets) with and without a leading-track selection of 5 \GeVc, in two \ptjet ranges: 10--15 \GeVc~ and 15--30 \GeVc. The fraction of gluon-initiated jets, calculated using PYTHIA 8 for with (without) the leading-track \pt selection, is  68\% (76\%) in the 10–-15 \GeVc~jet \pt interval and  75\% (76\%) in the 15–-30 \GeVc~jet \pt interval, indicating that these jets are gluon-dominated. Comparing the two inclusive-jet measurements shows the effect of requiring the leading hadron to have at least \pt > 5 \GeVc, which matches the \pt selection imposed on the \Dzero in the HF-tagged jets. For lower-\pt jets, this requirement modifies both the magnitude and peak location of the EEC, while for jets with \ptjet greater than 15 \GeVc, there is little effect.
The peak position is sensitive to the leading-particle \pt: as a result, 
it tends to shift toward the light-quark-initiated jet distribution. Notably, the leading-particle \pt selection is expected to impact the quark--gluon composition of the selected jets. The most striking feature in Fig.~\ref{fig:EECs_comp_pythia} is the significant difference in amplitude (peak value) between the charm-tagged and inclusive-jet EEC. These amplitude differences are quantified by the peak values, extracted by fitting the peak region of both charm-tagged and inclusive jets (without the leading-track \pt selection) with a {\texttt{Gaus(ln($R_{\rm L}$))}} function. This difference is $5\sigma$ and $3.5\sigma$ for the {\mbox{10--15 \GeVc}} and 15--30 \GeVc~ jet-\pt ranges, respectively. The significance is calculated by accounting for both statistical and systematic uncertainties of the data. This observation is consistent with the expected suppression of radiation from massive quarks due to the dead-cone effect~\cite{ALICE:deadcone}. This suppression is also evident in the middle panel, where the ratio of charm-tagged to inclusive-jet EEC highlights the differences in parton-shower evolution at large $R_{\rm L}$ and and shows suppression throughout the \Rl range, which becomes more pronounced at smaller angles. Despite the amplitude difference and its evolution, the peak positions of the charm-tagged and inclusive-jet EEC are similar, differing by only $1.6\sigma$ and $1\sigma$ compared to inclusive jets with and without the leading-track \pt-selection baseline in the 10--15 \GeVc~ jet-\pt range and showing negligible difference in the {\mbox{15--30 \GeVc~}} range. This similarity, despite the large mass of the charm quark compared to light quarks and gluons, suggests a complex interplay between flavor effects in the parton shower and a stronger contribution from non-perturbative effects such as hadronization~\cite{Larkoski:2017jix}. The flavor effects include the Casimir color factor (quarks vs. gluons), which leads to a broader distribution for inclusive jets, shifting the peak position to larger \Rl, and quark-mass effects (dead cone). 

These flavor effects are investigated with the help of the PYTHIA~8 Monash~\cite{Skands_2014} event generator. The EEC predicted by PYTHIA~8 for jets initiated by partons with different flavors is shown by solid lines in Fig.~\ref{fig:EECs_comp_pythia}, allowing for the isolation of quark-mass effects from Casimir color factors. PYTHIA~8 reproduces the observed peak amplitude of the charm-tagged jets EEC, which is also significantly lower than the PYTHIA~8 prediction for inclusive-jet EEC. The amplitude of charm-tagged jet EEC in data is lower than that predicted for light quark-initiated jets in PYTHIA~8 by 6.6$\sigma$ and 5$\sigma$ for the \ptjet ranges 10--15 and 15--30 \GeVc, respectively. The reduced amplitude for the charm-tagged jet EEC, compared to the light quark-initiated jet EEC, both of which are quark-initiated, further supports the expected suppression of radiation due to the dead-cone effect. The bottom panel, showing the ratio of charm-tagged to light-quark jets, reveals significantly more separation compared to the ratio with inclusive jets. PYTHIA~8 also reproduces minimal peak-position differences between the charm-tagged jet EEC and inclusive-jet EEC. Additionally, the simulations show that the differences in the peak positions of EEC between light quark-initiated jets in PYTHIA~8 and charm-tagged jets in data are 1.5$\sigma$ and 1.2$\sigma$ in the \ptjet ranges 10--15 and 15--30 \GeVc, respectively. This suggests that the differences in peak positions between charm-tagged jet EEC and both inclusive and light quark-initiated jet EEC should be small. This is intriguing because for low-\pt jets, mass effects such as the dead cone are expected to be dominant, determining the peak location difference~\cite{craft2022beautifulcharmingenergycorrelators}.

We compare our results to next-to-leading order (NLO) perturbative QCD calculations that include heavy-quark mass effects~\cite{craft2022beautifulcharmingenergycorrelators}, shown as the red band in Fig.~\ref{fig:EECs_comp_pythia}. The calculations reproduce the general shape of the measured EEC at large $R_{\rm L}$ and the lower amplitude for charm jets. 
However, assuming the uncertainty in the pQCD calculation is correlated in 
\Rl, the calculation predicts a peak position lower than that observed in the data, indicating that the omission of hadronization in the calculation impacts the ability to pinpoint the peak position. This discrepancy is more pronounced at low \ptjet. At very large $R_{\rm L}$, the slope of the pQCD calculations deviates from the data. This difference arises from the increasing edge effects as $R_{\rm L}$ approaches the jet radius, an effect absent in the pQCD calculations. These effects result from the drastic reduction in the number of pairs near the jet radius.

\begin{figure}[!ht]
    \centering
     \includegraphics[width=1\textwidth]{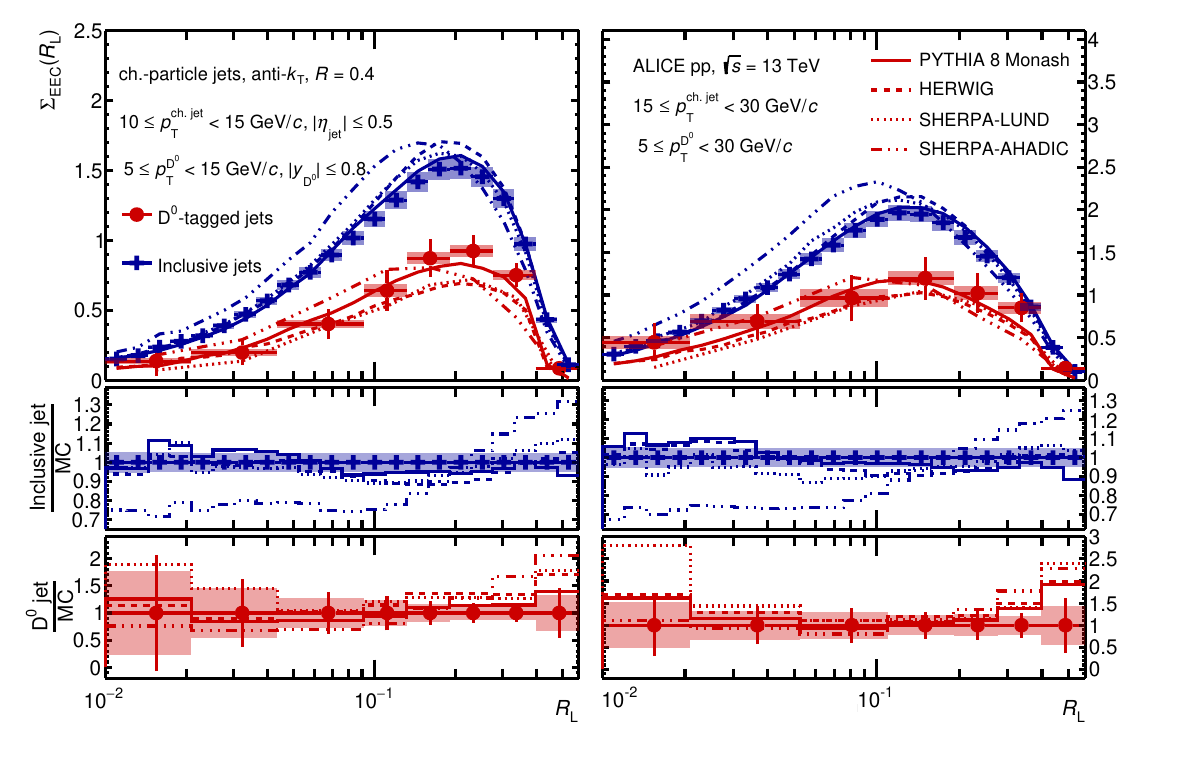}

    \caption{\Dzero-tagged jet (red) EEC distribution compared to inclusive jets without leading-track \pt selection (blue) in the \jetptrange{10}{15} (left) and \jetptrange{15}{30} (right) jet-\pt intervals. The data is shown in the solid points, and the different event generators (PYTHIA~8, Herwig, Sherpa Lund and Sherpa Ahadic) are shown in various line styles. The error bars show the statistical error, and the boxes represent the systematic error.}
    \label{fig:EECs_comp_mc_models}
\end{figure}

To further investigate the role of perturbative versus non-perturbative effects, such as hadronization, in Fig.~\ref{fig:EECs_comp_mc_models} we compare our data with MC predictions using different parton shower and hadronization models. This comparison allows us to identify the \Rl region most affected by hadronization and assess how much hadronization effects might bleed over into the perturbative region at large \Rl.  
PYTHIA~8~\cite{Skands_2014} uses the Lund string model, where hadrons are produced by breaking a color string into color-neutral hadrons and provides the best description of both charm-tagged and inclusive-jet EEC in the measured kinematic range. In contrast, HERWIG 7~\cite{Bellm:2015jjp} uses cluster hadronization, where locally color-connected partons are grouped into clusters that then decay into hadrons.  HERWIG 7 overpredicts the inclusive data, while underpredicting the charm-tagged jet data. This discrepancy suggests that the cluster-hadronization model may overproduce particles in the inclusive sample, whereas the underprediction for charm-tagged jets could arise from the treatment of reduced radiation from charm quarks due to the dead-cone effect, combined with subsequent hadronization effects. The agreement improves at higher~\ptjet.

Interpretation of the effect of the hadronization prescriptions is complicated by the fact that the parton-shower calculation in the two models is not identical. Therefore, we used the SHERPA 2.2.15~\cite{10.21468/SciPostPhys.7.3.034} model, where the parton shower and hadronization tunes can be adjusted independently, fixing 
the parton shower while varying only the hadronization tune.
Figure~\ref{fig:EECs_comp_mc_models}  shows that SHERPA-LUND~\cite{Sj_strand_2006}, which uses the Lund string hadronization model, tends to underpredict the data in the hadronization-dominated peak region for the charm-tagged jet EEC; at the same time, it overpredicts the inclusive-jet EEC. SHERPA-AHADIC, which uses the cluster-hadronization model, shows significant tension near the peak for both the charm-tagged jet EEC and the inclusive-jet EEC. The transition region (peak) from AHADIC occurs at a smaller angle compared to the LUND version, indicating that the cluster-hadronization model tends to cause a later hadronization compared to the hadronization by breaking an energetic ``string'' connecting the two originally struck partons. This highlights the sensitivity of the EEC observable to different hadronization mechanisms and indicates the need for improved theoretical modeling of jets, particularly in the transition region between perturbative and non-perturbative regimes.

In conclusion, we presented the first measurement of D$^0$-tagged jet EEC in pp collisions at {\mbox{$\sqrt{s} = 13$ TeV}} using the ALICE experiment. This novel observable provides qualitatively new insights into the dynamics of charm-quark jets, revealing the intricate interplay between perturbative and non-perturbative QCD processes~\cite{Chen:2020vvp, Komiske:2022enw}. Our key findings are twofold. First, the charm-tagged-jets EEC exhibit significantly lower amplitudes compared to inclusive jets, consistent with the expected suppression of radiation from massive partons due to the dead-cone effect~\cite{ALICE:deadcone}. This observation provides direct experimental evidence for the mass-dependent effects in parton-shower evolution. Second, the peak locations of the EEC distributions for charm-tagged and inclusive jets are remarkably similar. Theory expectations comparing light-quark and heavy-quark initiated showers, which are based solely on mass effects such as the dead cone, predict a peak position highly dependent on the heavy-quark mass~\cite{craft2022beautifulcharmingenergycorrelators}, which would be pronounced for low-\pt jets due to the larger dead-cone effect in this regime. Since inclusive jets are predominantly initiated by gluons, additional effects--such as the Casimir color factor in the parton shower and hadronization effects--also influence the EEC peak position within the reported kinematic range, resulting to a similar peak location.

The NLO pQCD calculations~\cite{craft2022beautifulcharmingenergycorrelators} reproduce the general features of our data but exhibit some tension, which is more pronounced for low-\pt jets. This discrepancy highlights the need for improved theoretical modeling of heavy-quark jets. Comparisons with various MC event generators, employing different prescriptions for parton showers and hadronization~\cite{Skands_2014, Bellm:2015jjp}, reveal the sensitivity of EECs to both flavor effects in the initial parton shower and the subsequent hadronization process. These results provide crucial constraints for refining the description of heavy-quark jets in QCD calculations and MC simulations.

Our measurements open new avenues for probing the fundamentals of QCD, offering a powerful tool to disentangle the effects of parton mass, color charge, and hadronization in jet evolution. Future studies extending this analysis to beauty-tagged jets and different collision systems will offer deeper insights into the dependence of jet substructure on the Casimir factor and quark-mass effects, as well as the properties of the quark--gluon plasma in heavy-ion collisions~\cite{Busza:2018rrf}.

\newenvironment{acknowledgement}{\relax}{\relax}
\begin{acknowledgement}
\section*{Acknowledgements}
We gratefully acknowledge Kyle Lee for providing theoretical predictions. 

The ALICE Collaboration would like to thank all its engineers and technicians for their invaluable contributions to the construction of the experiment and the CERN accelerator teams for the outstanding performance of the LHC complex.
The ALICE Collaboration gratefully acknowledges the resources and support provided by all Grid centres and the Worldwide LHC Computing Grid (WLCG) collaboration.
The ALICE Collaboration acknowledges the following funding agencies for their support in building and running the ALICE detector:
A. I. Alikhanyan National Science Laboratory (Yerevan Physics Institute) Foundation (ANSL), State Committee of Science and World Federation of Scientists (WFS), Armenia;
Austrian Academy of Sciences, Austrian Science Fund (FWF): [M 2467-N36] and Nationalstiftung f\"{u}r Forschung, Technologie und Entwicklung, Austria;
Ministry of Communications and High Technologies, National Nuclear Research Center, Azerbaijan;
Conselho Nacional de Desenvolvimento Cient\'{\i}fico e Tecnol\'{o}gico (CNPq), Financiadora de Estudos e Projetos (Finep), Funda\c{c}\~{a}o de Amparo \`{a} Pesquisa do Estado de S\~{a}o Paulo (FAPESP) and Universidade Federal do Rio Grande do Sul (UFRGS), Brazil;
Bulgarian Ministry of Education and Science, within the National Roadmap for Research Infrastructures 2020-2027 (object CERN), Bulgaria;
Ministry of Education of China (MOEC) , Ministry of Science \& Technology of China (MSTC) and National Natural Science Foundation of China (NSFC), China;
Ministry of Science and Education and Croatian Science Foundation, Croatia;
Centro de Aplicaciones Tecnol\'{o}gicas y Desarrollo Nuclear (CEADEN), Cubaenerg\'{\i}a, Cuba;
Ministry of Education, Youth and Sports of the Czech Republic, Czech Republic;
The Danish Council for Independent Research | Natural Sciences, the VILLUM FONDEN and Danish National Research Foundation (DNRF), Denmark;
Helsinki Institute of Physics (HIP), Finland;
Commissariat \`{a} l'Energie Atomique (CEA) and Institut National de Physique Nucl\'{e}aire et de Physique des Particules (IN2P3) and Centre National de la Recherche Scientifique (CNRS), France;
Bundesministerium f\"{u}r Bildung und Forschung (BMBF) and GSI Helmholtzzentrum f\"{u}r Schwerionenforschung GmbH, Germany;
General Secretariat for Research and Technology, Ministry of Education, Research and Religions, Greece;
National Research, Development and Innovation Office, Hungary;
Department of Atomic Energy Government of India (DAE), Department of Science and Technology, Government of India (DST), University Grants Commission, Government of India (UGC) and Council of Scientific and Industrial Research (CSIR), India;
National Research and Innovation Agency - BRIN, Indonesia;
Istituto Nazionale di Fisica Nucleare (INFN), Italy;
Japanese Ministry of Education, Culture, Sports, Science and Technology (MEXT) and Japan Society for the Promotion of Science (JSPS) KAKENHI, Japan;
Consejo Nacional de Ciencia (CONACYT) y Tecnolog\'{i}a, through Fondo de Cooperaci\'{o}n Internacional en Ciencia y Tecnolog\'{i}a (FONCICYT) and Direcci\'{o}n General de Asuntos del Personal Academico (DGAPA), Mexico;
Nederlandse Organisatie voor Wetenschappelijk Onderzoek (NWO), Netherlands;
The Research Council of Norway, Norway;
Pontificia Universidad Cat\'{o}lica del Per\'{u}, Peru;
Ministry of Science and Higher Education, National Science Centre and WUT ID-UB, Poland;
Korea Institute of Science and Technology Information and National Research Foundation of Korea (NRF), Republic of Korea;
Ministry of Education and Scientific Research, Institute of Atomic Physics, Ministry of Research and Innovation and Institute of Atomic Physics and Universitatea Nationala de Stiinta si Tehnologie Politehnica Bucuresti, Romania;
Ministerstvo skolstva, vyskumu, vyvoja a mladeze SR, Slovakia;
National Research Foundation of South Africa, South Africa;
Swedish Research Council (VR) and Knut \& Alice Wallenberg Foundation (KAW), Sweden;
European Organization for Nuclear Research, Switzerland;
Suranaree University of Technology (SUT), National Science and Technology Development Agency (NSTDA) and National Science, Research and Innovation Fund (NSRF via PMU-B B05F650021), Thailand;
Turkish Energy, Nuclear and Mineral Research Agency (TENMAK), Turkey;
National Academy of  Sciences of Ukraine, Ukraine;
Science and Technology Facilities Council (STFC), United Kingdom;
National Science Foundation of the United States of America (NSF) and United States Department of Energy, Office of Nuclear Physics (DOE NP), United States of America.
In addition, individual groups or members have received support from:
Czech Science Foundation (grant no. 23-07499S), Czech Republic;
FORTE project, reg.\ no.\ CZ.02.01.01/00/22\_008/0004632, Czech Republic, co-funded by the European Union, Czech Republic;
European Research Council (grant no. 950692), European Union;
Deutsche Forschungs Gemeinschaft (DFG, German Research Foundation) ``Neutrinos and Dark Matter in Astro- and Particle Physics'' (grant no. SFB 1258), Germany;
ICSC - National Research Center for High Performance Computing, Big Data and Quantum Computing and FAIR - Future Artificial Intelligence Research, funded by the NextGenerationEU program (Italy).

\end{acknowledgement}

\bibliographystyle{utphys}   
\bibliography{main}

\newpage
\appendix
\section{The ALICE Collaboration}
\label{app:collab}
\begin{flushleft} 
\small

S.~Acharya\,\orcidlink{0000-0002-9213-5329}\,$^{\rm 50}$, 
A.~Agarwal$^{\rm 133}$, 
G.~Aglieri Rinella\,\orcidlink{0000-0002-9611-3696}\,$^{\rm 32}$, 
L.~Aglietta\,\orcidlink{0009-0003-0763-6802}\,$^{\rm 24}$, 
M.~Agnello\,\orcidlink{0000-0002-0760-5075}\,$^{\rm 29}$, 
N.~Agrawal\,\orcidlink{0000-0003-0348-9836}\,$^{\rm 25}$, 
Z.~Ahammed\,\orcidlink{0000-0001-5241-7412}\,$^{\rm 133}$, 
S.~Ahmad\,\orcidlink{0000-0003-0497-5705}\,$^{\rm 15}$, 
S.U.~Ahn\,\orcidlink{0000-0001-8847-489X}\,$^{\rm 71}$, 
I.~Ahuja\,\orcidlink{0000-0002-4417-1392}\,$^{\rm 36}$
% ...

\section*{Affiliation Notes}

$^{\rm I}$ Deceased\\
$^{\rm II}$ Also at: Max-Planck-Institut fur Physik, Munich, Germany\\
$^{\rm III}$ Also at: Italian National Agency for New Technologies, Energy and Sustainable Economic Development (ENEA), Bologna, Italy\\
$^{\rm IV}$ Also at: Instituto de Fisica da Universidade de Sao Paulo\\
$^{\rm V}$ Also at: Dipartimento DET del Politecnico di Torino, Turin, Italy\\
$^{\rm VI}$ Also at: Department of Applied Physics, Aligarh Muslim University, Aligarh, India\\
$^{\rm VII}$ Also at: Institute of Theoretical Physics, University of Wroclaw, Poland\\
$^{\rm VIII}$ Also at: Facultad de Ciencias, Universidad Nacional Aut\'{o}noma de M\'{e}xico, Mexico City, Mexico\\

\section*{Collaboration Institutes}

$^{1}$ A.I. Alikhanyan National Science Laboratory (Yerevan Physics Institute) Foundation, Yerevan, Armenia\\
$^{2}$ AGH University of Krakow, Cracow, Poland\\
$^{3}$ Bogolyubov Institute for Theoretical Physics, National Academy of Sciences of Ukraine, Kiev, Ukraine\\
$^{4}$ Bose Institute, Department of Physics  and Centre for Astroparticle Physics and Space Science (CAPSS), Kolkata, India\\
$^{5}$ California Polytechnic State University, San Luis Obispo, California, United States\\
$^{6}$ Central China Normal University, Wuhan, China\\
$^{7}$ Centro de Aplicaciones Tecnol\'{o}gicas y Desarrollo Nuclear (CEADEN), Havana, Cuba\\
$^{8}$ Centro de Investigaci\'{o}n y de Estudios Avanzados (CINVESTAV), Mexico City and M\'{e}rida, Mexico\\
$^{9}$ Chicago State University, Chicago, Illinois, United States\\
$^{10}$ China Nuclear Data Center, China Institute of Atomic Energy, Beijing, China\\
$^{11}$ China University of Geosciences, Wuhan, China\\
$^{12}$ Chungbuk National University, Cheongju, Republic of Korea\\
$^{13}$ Comenius University Bratislava, Faculty of Mathematics, Physics and Informatics, Bratislava, Slovak Republic\\
$^{14}$ Creighton University, Omaha, Nebraska, United States\\
$^{15}$ Department of Physics, Aligarh Muslim University, Aligarh, India\\
$^{16}$ Department of Physics, Pusan National University, Pusan, Republic of Korea\\
$^{17}$ Department of Physics, Sejong University, Seoul, Republic of Korea\\
$^{18}$ Department of Physics, University of California, Berkeley, California, United States\\
$^{19}$ Department of Physics, University of Oslo, Oslo, Norway\\
$^{20}$ Department of Physics and Technology, University of Bergen, Bergen, Norway\\
$^{21}$ Dipartimento di Fisica, Universit\`{a} di Pavia, Pavia, Italy\\
$^{22}$ Dipartimento di Fisica dell'Universit\`{a} and Sezione INFN, Cagliari, Italy\\
$^{23}$ Dipartimento di Fisica dell'Universit\`{a} and Sezione INFN, Trieste, Italy\\
$^{24}$ Dipartimento di Fisica dell'Universit\`{a} and Sezione INFN, Turin, Italy\\
$^{25}$ Dipartimento di Fisica e Astronomia dell'Universit\`{a} and Sezione INFN, Bologna, Italy\\
$^{26}$ Dipartimento di Fisica e Astronomia dell'Universit\`{a} and Sezione INFN, Catania, Italy\\
$^{27}$ Dipartimento di Fisica e Astronomia dell'Universit\`{a} and Sezione INFN, Padova, Italy\\
$^{28}$ Dipartimento di Fisica `E.R.~Caianiello' dell'Universit\`{a} and Gruppo Collegato INFN, Salerno, Italy\\
$^{29}$ Dipartimento DISAT del Politecnico and Sezione INFN, Turin, Italy\\
$^{30}$ Dipartimento di Scienze MIFT, Universit\`{a} di Messina, Messina, Italy\\
$^{31}$ Dipartimento Interateneo di Fisica `M.~Merlin' and Sezione INFN, Bari, Italy\\
$^{32}$ European Organization for Nuclear Research (CERN), Geneva, Switzerland\\
$^{33}$ Faculty of Electrical Engineering, Mechanical Engineering and Naval Architecture, University of Split, Split, Croatia\\
$^{34}$ Faculty of Nuclear Sciences and Physical Engineering, Czech Technical University in Prague, Prague, Czech Republic\\
$^{35}$ Faculty of Physics, Sofia University, Sofia, Bulgaria\\
$^{36}$ Faculty of Science, P.J.~\v{S}af\'{a}rik University, Ko\v{s}ice, Slovak Republic\\
$^{37}$ Faculty of Technology, Environmental and Social Sciences, Bergen, Norway\\
$^{38}$ Frankfurt Institute for Advanced Studies, Johann Wolfgang Goethe-Universit\"{a}t Frankfurt, Frankfurt, Germany\\
$^{39}$ Fudan University, Shanghai, China\\
$^{40}$ Gangneung-Wonju National University, Gangneung, Republic of Korea\\
$^{41}$ Gauhati University, Department of Physics, Guwahati, India\\
$^{42}$ Helmholtz-Institut f\"{u}r Strahlen- und Kernphysik, Rheinische Friedrich-Wilhelms-Universit\"{a}t Bonn, Bonn, Germany\\
$^{43}$ Helsinki Institute of Physics (HIP), Helsinki, Finland\\
$^{44}$ High Energy Physics Group,  Universidad Aut\'{o}noma de Puebla, Puebla, Mexico\\
$^{45}$ Horia Hulubei National Institute of Physics and Nuclear Engineering, Bucharest, Romania\\
$^{46}$ HUN-REN Wigner Research Centre for Physics, Budapest, Hungary\\
$^{47}$ Indian Institute of Technology Bombay (IIT), Mumbai, India\\
$^{48}$ Indian Institute of Technology Indore, Indore, India\\
$^{49}$ INFN, Laboratori Nazionali di Frascati, Frascati, Italy\\
$^{50}$ INFN, Sezione di Bari, Bari, Italy\\
$^{51}$ INFN, Sezione di Bologna, Bologna, Italy\\
$^{52}$ INFN, Sezione di Cagliari, Cagliari, Italy\\
$^{53}$ INFN, Sezione di Catania, Catania, Italy\\
$^{54}$ INFN, Sezione di Padova, Padova, Italy\\
$^{55}$ INFN, Sezione di Pavia, Pavia, Italy\\
$^{56}$ INFN, Sezione di Torino, Turin, Italy\\
$^{57}$ INFN, Sezione di Trieste, Trieste, Italy\\
$^{58}$ Inha University, Incheon, Republic of Korea\\
$^{59}$ Institute for Gravitational and Subatomic Physics (GRASP), Utrecht University/Nikhef, Utrecht, Netherlands\\
$^{60}$ Institute of Experimental Physics, Slovak Academy of Sciences, Ko\v{s}ice, Slovak Republic\\
$^{61}$ Institute of Physics, Homi Bhabha National Institute, Bhubaneswar, India\\
$^{62}$ Institute of Physics of the Czech Academy of Sciences, Prague, Czech Republic\\
$^{63}$ Institute of Space Science (ISS), Bucharest, Romania\\
$^{64}$ Institut f\"{u}r Kernphysik, Johann Wolfgang Goethe-Universit\"{a}t Frankfurt, Frankfurt, Germany\\
$^{65}$ Instituto de Ciencias Nucleares, Universidad Nacional Aut\'{o}noma de M\'{e}xico, Mexico City, Mexico\\
$^{66}$ Instituto de F\'{i}sica, Universidade Federal do Rio Grande do Sul (UFRGS), Porto Alegre, Brazil\\
$^{67}$ Instituto de F\'{\i}sica, Universidad Nacional Aut\'{o}noma de M\'{e}xico, Mexico City, Mexico\\
$^{68}$ iThemba LABS, National Research Foundation, Somerset West, South Africa\\
$^{69}$ Jeonbuk National University, Jeonju, Republic of Korea\\
$^{70}$ Johann-Wolfgang-Goethe Universit\"{a}t Frankfurt Institut f\"{u}r Informatik, Fachbereich Informatik und Mathematik, Frankfurt, Germany\\
$^{71}$ Korea Institute of Science and Technology Information, Daejeon, Republic of Korea\\
$^{72}$ Laboratoire de Physique Subatomique et de Cosmologie, Universit\'{e} Grenoble-Alpes, CNRS-IN2P3, Grenoble, France\\
$^{73}$ Lawrence Berkeley National Laboratory, Berkeley, California, United States\\
$^{74}$ Lund University Department of Physics, Division of Particle Physics, Lund, Sweden\\
$^{75}$ Nagasaki Institute of Applied Science, Nagasaki, Japan\\
$^{76}$ Nara Women{'}s University (NWU), Nara, Japan\\
$^{77}$ National and Kapodistrian University of Athens, School of Science, Department of Physics , Athens, Greece\\
$^{78}$ National Centre for Nuclear Research, Warsaw, Poland\\
$^{79}$ National Institute of Science Education and Research, Homi Bhabha National Institute, Jatni, India\\
$^{80}$ National Nuclear Research Center, Baku, Azerbaijan\\
$^{81}$ National Research and Innovation Agency - BRIN, Jakarta, Indonesia\\
$^{82}$ Niels Bohr Institute, University of Copenhagen, Copenhagen, Denmark\\
$^{83}$ Nikhef, National institute for subatomic physics, Amsterdam, Netherlands\\
$^{84}$ Nuclear Physics Group, STFC Daresbury Laboratory, Daresbury, United Kingdom\\
$^{85}$ Nuclear Physics Institute of the Czech Academy of Sciences, Husinec-\v{R}e\v{z}, Czech Republic\\
$^{86}$ Oak Ridge National Laboratory, Oak Ridge, Tennessee, United States\\
$^{87}$ Ohio State University, Columbus, Ohio, United States\\
$^{88}$ Physics department, Faculty of science, University of Zagreb, Zagreb, Croatia\\
$^{89}$ Physics Department, Panjab University, Chandigarh, India\\
$^{90}$ Physics Department, University of Jammu, Jammu, India\\
$^{91}$ Physics Program and International Institute for Sustainability with Knotted Chiral Meta Matter (WPI-SKCM$^{2}$), Hiroshima University, Hiroshima, Japan\\
$^{92}$ Physikalisches Institut, Eberhard-Karls-Universit\"{a}t T\"{u}bingen, T\"{u}bingen, Germany\\
$^{93}$ Physikalisches Institut, Ruprecht-Karls-Universit\"{a}t Heidelberg, Heidelberg, Germany\\
$^{94}$ Physik Department, Technische Universit\"{a}t M\"{u}nchen, Munich, Germany\\
$^{95}$ Politecnico di Bari and Sezione INFN, Bari, Italy\\
$^{96}$ Research Division and ExtreMe Matter Institute EMMI, GSI Helmholtzzentrum f\"ur Schwerionenforschung GmbH, Darmstadt, Germany\\
$^{97}$ Saga University, Saga, Japan\\
$^{98}$ Saha Institute of Nuclear Physics, Homi Bhabha National Institute, Kolkata, India\\
$^{99}$ School of Physics and Astronomy, University of Birmingham, Birmingham, United Kingdom\\
$^{100}$ Secci\'{o}n F\'{\i}sica, Departamento de Ciencias, Pontificia Universidad Cat\'{o}lica del Per\'{u}, Lima, Peru\\
$^{101}$ Stefan Meyer Institut f\"{u}r Subatomare Physik (SMI), Vienna, Austria\\
$^{102}$ SUBATECH, IMT Atlantique, Nantes Universit\'{e}, CNRS-IN2P3, Nantes, France\\
$^{103}$ Sungkyunkwan University, Suwon City, Republic of Korea\\
$^{104}$ Suranaree University of Technology, Nakhon Ratchasima, Thailand\\
$^{105}$ Technical University of Ko\v{s}ice, Ko\v{s}ice, Slovak Republic\\
$^{106}$ The Henryk Niewodniczanski Institute of Nuclear Physics, Polish Academy of Sciences, Cracow, Poland\\
$^{107}$ The University of Texas at Austin, Austin, Texas, United States\\
$^{108}$ Universidad Aut\'{o}noma de Sinaloa, Culiac\'{a}n, Mexico\\
$^{109}$ Universidade de S\~{a}o Paulo (USP), S\~{a}o Paulo, Brazil\\
$^{110}$ Universidade Estadual de Campinas (UNICAMP), Campinas, Brazil\\
$^{111}$ Universidade Federal do ABC, Santo Andre, Brazil\\
$^{112}$ Universitatea Nationala de Stiinta si Tehnologie Politehnica Bucuresti, Bucharest, Romania\\
$^{113}$ University of Derby, Derby, United Kingdom\\
$^{114}$ University of Houston, Houston, Texas, United States\\
$^{115}$ University of Jyv\"{a}skyl\"{a}, Jyv\"{a}skyl\"{a}, Finland\\
$^{116}$ University of Kansas, Lawrence, Kansas, United States\\
$^{117}$ University of Liverpool, Liverpool, United Kingdom\\
$^{118}$ University of Science and Technology of China, Hefei, China\\
$^{119}$ University of South-Eastern Norway, Kongsberg, Norway\\
$^{120}$ University of Tennessee, Knoxville, Tennessee, United States\\
$^{121}$ University of the Witwatersrand, Johannesburg, South Africa\\
$^{122}$ University of Tokyo, Tokyo, Japan\\
$^{123}$ University of Tsukuba, Tsukuba, Japan\\
$^{124}$ Universit\"{a}t M\"{u}nster, Institut f\"{u}r Kernphysik, M\"{u}nster, Germany\\
$^{125}$ Universit\'{e} Clermont Auvergne, CNRS/IN2P3, LPC, Clermont-Ferrand, France\\
$^{126}$ Universit\'{e} de Lyon, CNRS/IN2P3, Institut de Physique des 2 Infinis de Lyon, Lyon, France\\
$^{127}$ Universit\'{e} de Strasbourg, CNRS, IPHC UMR 7178, F-67000 Strasbourg, France, Strasbourg, France\\
$^{128}$ Universit\'{e} Paris-Saclay, Centre d'Etudes de Saclay (CEA), IRFU, D\'{e}partment de Physique Nucl\'{e}aire (DPhN), Saclay, France\\
$^{129}$ Universit\'{e}  Paris-Saclay, CNRS/IN2P3, IJCLab, Orsay, France\\
$^{130}$ Universit\`{a} degli Studi di Foggia, Foggia, Italy\\
$^{131}$ Universit\`{a} del Piemonte Orientale, Vercelli, Italy\\
$^{132}$ Universit\`{a} di Brescia, Brescia, Italy\\
$^{133}$ Variable Energy Cyclotron Centre, Homi Bhabha National Institute, Kolkata, India\\
$^{134}$ Warsaw University of Technology, Warsaw, Poland\\
$^{135}$ Wayne State University, Detroit, Michigan, United States\\
$^{136}$ Yale University, New Haven, Connecticut, United States\\
$^{137}$ Yildiz Technical University, Istanbul, Turkey\\
$^{138}$ Yonsei University, Seoul, Republic of Korea\\
$^{139}$ Affiliated with an institute formerly covered by a cooperation agreement with CERN\\
$^{140}$ Affiliated with an international laboratory covered by a cooperation agreement with CERN.\\

\end{flushleft} 
\end{document}